\documentclass[12pt]{article}
\usepackage[ansinew]{inputenc}
\usepackage{array}
\usepackage{color}
\usepackage{lscape}
\usepackage{amsmath}
\usepackage{amsxtra}
\usepackage{amstext}
\usepackage{amssymb}
\usepackage{latexsym}
\usepackage{dsfont}
\usepackage{graphicx}
\usepackage{graphics}
\usepackage{epstopdf}
\DeclareGraphicsExtensions{.eps,.ps,.eps.gz,.ps.gz,.eps.Z}

\begin{document}
\markright{\underline{Pricing European Options by Stable Fourier-Cosine Series Expansions}}
\title{\textbf{Pricing European Options by Stable Fourier-Cosine Series Expansions}}
\author{Chunfa Wang\\
Fiance School of Zhejiang University of Finance and Economics,\\
Hangzhou, China, cfwang@zufe.edu.cn}
\date{January 2, 2017}
\maketitle

\begin{abstract}
The COS method proposed in Fang and Oosterlee (2008), although highly efficient, may lack robustness for a number of cases.
In this paper, we present a Stable pricing of call options based on Fourier cosine series expansion.
The Stability of the pricing methods is demonstrated by error analysis, as
well as by a series of numerical examples, including the Heston stochastic volatility model, Kou jump-diffusion model,
and CGMY model.
\end{abstract}

\section{Introduction}

A fundamental problem of option pricing is the explicit computation of discounted expected value
which arise as prices of derivatives. Efficient methods to compute such expectations are crucial in particular for
calibration purposes. During a calibration procedure in each iteration step typically
a large number of model prices has to be computed and compared to market prices.
Therefore, a fast yet accurate compute method is demanded.

A method which almost always works to get expectations
is Monte Carlo simulation. Its disadvantage is that it is computer intensive
and therefore too slow for many purposes. Another classical approach is to represent
prices as solutions of partial (integro-) differential equations (PDEs.
This approach applies to a wide range of valuation problems, in particular it allows
to compute prices of American options as well. Nevertheless the numerical solution
of PIDEs rests on sophisticated discretization methods and corresponding programs.
A third approach is numerical integration methods. The latter type of methods
is attractive from both practice and research point of view, as the fast computational
speed, especially for plain vanilla options, makes it useful for calibration at financial institutions.

Usually numerical integration techniques are combined with the
Fourier transform or Hilbert transform, and therefore, the numerical
integration methods are often referred to as the ``transform methods".
The initial references for Fourier transform methods to compute option prices are
Carr and Madan (1999) and Raible (2000). Whereas the first mentioned authors
consider Fourier transforms of appropriately modified call prices and then invert
these, the second author starts with representing the option price as a convolution
of the modified payoff and the log return density, then derives the bilateral Laplace
transform and finally inverts the resulting product. In both cases the result is an
integral which can be evaluated numerically fast.

A recent contribution to the transform method category is the COS
method proposed in Fang and Oosterlee (2008)---a numerical approximation
based on the Fourier cosine series expansion. Fang and Oosterlee (2008) show that
the convergence rate for this method is exponential with linear computational
complexity in most cases. The method was then used to price early-exercise
and discrete barrier options in Fang and Oosterlee (2009), Asian options in
Zhang and Oosterlee (2013), and Bermudan options in the Heston model in Fang and Oosterlee (2011).

As Fang and Oosterlee (2008) and Zhang and Oosterlee (2011) pointed:
When pricing call options with the COS method, the method's accuracy
may exhibit sensitivity regarding the choice of the domain size in which
the series expansion is defined. A call payoff grows exponentially with the
log-stock price which may introduce significant cancellation errors for large
domain sizes. Put options do not suffer from this, as their payoff value is
bounded by the strike value. For pricing European calls, one can employ the
well-known put-call parity or put-call duality and price calls via puts.

In this paper, we present a stable pricing of call options based on Fourier cosine series expansion.
Since the conditional probability density function $f(y|x)$ of the underlying decays to zero rapidly as $y\rightarrow\pm\infty$,
$e^{\alpha y}f(y|x)$ still decays to zero rapidly for appropriate values $\alpha$.
We take Fourier cosine series expansion for $e^{\alpha y}f(y|x)$ which allows us damping payoff function of option by a factor $e^{-\alpha y}$.
Therefore the growth rate of $e^{-\alpha y}g(y)$ is decreased when $\alpha>0$ and the cancellation error for large values of $L$ is reduced.
The robustness of the pricing methods is demonstrated by error analysis, as
well as by a series of numerical examples, including the Heston stochastic volatility model, Kou jump-diffusion model
and CGMY model.

The outline of the paper is as follows: In Section 2 we present the option pricing problem
and explain stable Cos methods for the option pricing problem.
The error analysis is also presented in this section. Section 3 then presents a variety of numerical results,
confirming our robust version of the COS valuation method. Finally, Section 6 is devoted to
conclusions.

\section{Stable Cos methods for Pricing European Call Option}

Let $(\Omega,\mathcal{F}, \{\mathcal{F}_{t}\}_{0\leq t\leq T}, P)$ be a filtered probability space,
where $P$ is a risk neutral measure, $\mathbb{F} =\{\mathcal{F}_{t}\}_{0\leq t\leq T}$ satisfies the usual
hypotheses of completeness and right continuity, $T > 0$ a finite terminal time.
The asset price process $\{S_{t}\}_{0\leq t \leq T}$ is a stochastic
process on the filtered probability space $(\Omega,\mathcal{F}, \mathbb{F}, P)$.
Let us consider a European type claim whose payoff at maturity $T$ is given by $g(Y_{T})$,
where $g(\cdot)$ is a function on $\mathbb{R}$,
$Y_{t}=\log (S_{t}/K)$ and $K$ is the strike price.
The value of such claim at time 0 is given by the risk-neutral option valuation formula
\begin{equation}\label{1}
    v(0,x)=e^{-rT}E[g(Y_{T})|x]=e^{-rT}\int^{\infty}_{-\infty}g(y)f_{T}(y|x)dy,
\end{equation}
where $x=Y_{0}$ is the current state, $f_{T}(y|x)$ is the conditional density function,
$r$ is the risk-free rate. We assume that the characteristic function of $\{Y_{t}\}_{0\leq t \leq T}$ is known,
which is the usually case, and the integrand is integrable, which is common for most problems we deal
with.

First, for given $x$, we truncate the infinite integration ranges to some interval
$[a,b]\subset\mathbb{R}$ without loosing significant accuracy and obtain approximation $v_{1}$
\begin{equation}\label{2}
v(0,x)\approx v_{1}(0,x)\triangleq e^{-rT}\int^{b}_{a}g(y)f_{T}(y|x)dy
\end{equation}
As Fang and Oosterlee (2008), $[a,b]$ can be taken $[a,b]$ as
\begin{equation}\label{3}
\begin{split}
a=&c_{1}[Y_{T}]-L\sqrt{c_{2}[Y_{T}]+\sqrt{c_{4}[Y_{T}]}}\\
b=&c_{1}[Y_{T}]+L\sqrt{c_{2}[Y_{T}]+\sqrt{c_{4}[Y_{T}]}}
\end{split}
\end{equation}
where $c_{n}[Y_{T}]$ denotes the $n$-th cumulant of $Y_{T}$.

\subsection{Fang-Oosterlee Cos method}

In Fang-Oosterlee Cos method, the conditional density function is approximated on a truncated domain, by a
truncated Fourier cosine expansion, which recovers the conditional density function
from its characteristic function as follows:
\begin{equation}\label{4}
f_{T}(y|x)\approx\frac{2}{b-a}\sum^{N-1}_{k=0}{'}\mathrm{Re}\bigg[\varphi_{T}\bigg(\frac{k\pi}{b-a},x\bigg)
\exp\bigg(-ik\pi\frac{a}{b-a}\bigg)\bigg]\cos\bigg(k\pi\frac{y-a}{b-a}\bigg),
\end{equation}
with $\varphi_{T}(u,x)$ the characteristic function of $f_{T}(y|x)$ and $\mathrm{Re}[\cdot]$ means taking the real part of the argument.
The $\sum'$ indicates that the first term in the summation is weighted by one-half.

Replacing $f_{T}(y|x)$ by its approximation (4) in Equation (3) and interchanging integration
and summation gives the COS formula for computing the values of European
options:
\begin{equation}\label{5}
v(0,x) = e^{-r\Delta t}
\sum^{N-1}_{k=0}{'}\mathrm{Re}\bigg[\varphi_{T}\bigg(\frac{k\pi}{b-a},x\bigg)\exp\bigg(-ik\pi\frac{a}{b-a}\bigg)\bigg]V_{k},
\end{equation}
where:
\begin{equation*}
V_{k} =\frac{2}{b-a}\int^{b}_{a}g(y)\cos\bigg(k\pi\frac{y-a}{b-a}\bigg)dy,
\end{equation*}
are the Fourier cosine coefficients of $g(y)$, that are available in closed form for
several payoff functions, like for plain vanilla puts and calls, but also for example
for discontinuous payoffs like for digital options.

It was shown in Fang and Oosterlee (2008), that, with integration interval $[a,b]$
chosen sufficiently wide, the series truncation error dominates the overall error. For
conditional density functions $f_{T}(y|x)\in C^{\infty}((a,b)\subset\mathbb{R}$), the method converges exponentially;
otherwise convergence is algebraically.

However, when pricing call options, the solution's accuracy exhibits sensitivity regarding
the size of this truncated domain. This holds specifically for call options
under fat-tailed distributions, like under certain L\'{e}vy jump processes, or for options
with a very long time to maturity.\footnote{This is mainly the case when we consider real options
or insurance products with a long life time.} A call payoff grows exponentially in log-stock
price which may introduce cancellation errors for large domain sizes. A put option
does not suffer from this (see Fang and Oosterlee (2009)), as their payoff value is bounded by the strike
value. In Fang and Oosterlee (2008), European call options were therefore priced by means of European
put option computations, in combination with the put-call parity:
\begin{equation}\label{6}
v^{\mathrm{call}}(0,x) = v^{\mathrm{put}}(0,x)+S_{t}e^{-qT)}-Ke^{-rT},
\end{equation}
where $v^{\mathrm{call}}(0,x)$ and $v^{\mathrm{put}}(0,x)$ are the call and put option prices, respectively, and $q$
is again the dividend rate. The parity lead to robust formulas for pricing European call options by the
COS method.

\subsection{Stable Cos method}

In this section, we present a robust pricing of European call options by Fourier-cosine series expansion.
Since the density $f_{T}(y|x)$ decays to zero rapidly as $y\rightarrow\pm\infty$, we first modify the density $f_{T}(y|x)$ by multiplying a factor
$e^{\alpha y}$, then take Fourier-cosine expansion for $e^{\alpha y}f_{T}(y|x)$ which reads as
\begin{equation}\label{7}
e^{\alpha y}f_{T}(y|x)=\sum^{\infty}_{n=0}{^{'}}A_{T}(u_{n},x)\cos[u_{n}(y-a)]:=\widetilde{f}_{T}(y|x)
\end{equation}
where $\alpha\in\mathbb{R}$, $u_{n}=n\pi/(b-a)$, and
\begin{equation}\label{8}
A_{T}(u,x)=\frac{2}{b-a}\int^{b}_{a}e^{\alpha y}f_{T}(y|x)\cos[u(y-a)]dy.
\end{equation}
Then replace the density $f_{T}(y|x)$ by $e^{-\alpha y}\widetilde{f}_{T}(y|x)$ in (2), so we obtain
\begin{equation*}
v_{1}(0,x)= e^{-rT}\int^{b}_{a}e^{-\alpha y}g(y)\sum^{\infty}_{n=0}{^{'}}A_{T}(u_{n},x)\cos[u_{n}(y-a)]dy
\end{equation*}
We interchange the summation and integration, and insert the define
\begin{equation}\label{9}
V_{T}(u)\equiv\frac{2}{b-a}\int^{b}_{a}e^{-\alpha y}g(y)\cos[u(y-a)]dy
\end{equation}
resulting
\begin{equation}\label{10}
v_{1}(0,x)= \frac{b-a}{2}e^{-rT}\sum^{\infty}_{n=0}{^{'}}A_{T}(u_{n},x)V_{T}(u_{n})
\end{equation}

\textbf{Remark 1.} When payoff function $g(y)$ grows exponentially, we can choose $\alpha>0$ such that the growth rate of $e^{-\alpha y}g(y)$
is decreased and therefore the cancellation error for large values of $L$ is reduced. $\alpha$ can thus be seen as a damping factor.

\vspace{2mm}

Next, we truncate the series summation, resulting in approximation $v_{3}$
\begin{equation}\label{11}
v_{1}(0,x)\approx v_{2}(0,x)\triangleq \frac{b-a}{2}e^{-rT}\sum^{N-1}_{n=0}{^{'}}A_{T}(u_{n},x)V_{T}(u_{n})
\end{equation}
Finally, same as Fang and Oosterlee (2008), for $u\in\mathbb{R}$, the coefficients $A_{T}(u,x)$ are approximated by
\begin{align}\label{12}
\overline{A}_{T}(u,x)=&\frac{2}{b-a}\int^{\infty}_{-\infty}e^{\alpha y}f_{T}(y|x)\cos[u(y-a)]dy\nonumber\\
=&\frac{2}{b-a}\mathrm{Re}\big[e^{-iua}\widetilde{\phi}_{T}(u-i\alpha)\big]
\end{align}
where $\widetilde{\phi}_{T}(\cdot)$ is the conditional
characteristic function of $Y_{T}$, given $Y_{0}= x$.
Denotes $X_{T}=Y_{T}-x$, and $\phi_{T}(u)$ the characteristic function of $X_{T}$.
Then $\widetilde{\phi}_{T}(u)=e^{iux}\phi_{T}(u)$. Thus
\begin{align}\label{13}
\overline{A}_{T}(u,x)=&\frac{2}{b-a}\mathrm{Re}\bigg[e^{-iua}e^{i(u-i\alpha)x}\phi_{T}(u-i\alpha)\bigg]\nonumber\\
=&\frac{2e^{\alpha x}}{b-a}\mathrm{Re}\bigg[e^{iu(x-a)}\phi_{T}(u-i\alpha)\bigg]
\end{align}
Replacing $A_{T}(u,x)$ by $\overline{A}_{T}(u,x)$ in (11), we obtain
\begin{equation}\label{14}
v_{2}(0,x)\approx v_{3}(0,x)\triangleq
\frac{b-a}{2}e^{-rT}\sum^{N-1}_{n=0}{^{'}}\overline{A}_{T}(u_{n},x)V_{T}(u_{n})
\end{equation}

\subsection{Error Analysis}

In this subsection we give error analysis for the stable COS pricing method. First, we analyze the local error,
i.e., the error in the continuation values at each time step. A similar error analysis has been performed in [13],
where, however, the influence of the call payoff function on the global error convergence
was omitted. Here, we study the influence of the payoff function and the integration range on the error convergence.

It has been shown, in Fang and Oosterlee (2008), that the error of the COS method for the error in the continuation
value consists of three parts, denoted by $\varepsilon_{1}$, $\varepsilon_{2}$ and $\varepsilon_{3}$, respectively.

Error $\varepsilon_{1}$ is the integration range error
\begin{equation*}
|\varepsilon_{1}(x, [a,b])| = e^{-rT} \int_{\mathbb{R}\backslash[a,b]}g(y)f_{T}(y|x)dy,
\end{equation*}
which depends on the payoff function and the integration range.

Error $\varepsilon_{2}$ is the series truncation error on $[a,b]$, which depends on the smoothness
of the probability density function of the underlying processes:
\begin{equation}\label{15}
\varepsilon_{2}(x;N, [a,b]):=e^{-rT}\sum^{\infty}_{k=N}\mathrm{Re}\bigg[e^{-ik\pi\frac{a}{b-a}}
\int^{b}_{a}e^{ik\pi\frac{y}{b-a}}e^{\alpha y}f_{T}(y|x)dy\bigg]V_{k}.
\end{equation}
For probability density functions $f_{T}(y|x)\subset C^{\infty}[a,b]$, we have
\begin{equation*}
|\varepsilon_{2}(x,N,[a,b])| < P\exp(-(N -1)\nu),
\end{equation*}
where $N$ is the number of terms in the Fourier cosine expansions, $\nu >0$ is a constant
and $P$ is a term which varies less than exponentially with respect to $N$. When the
probability density function has a discontinuous derivative, then the Fourier cosine
expansions converge algebraically,
\begin{equation*}
|\varepsilon_{2}(x,N,[a,b])| <\frac{P}{(N-1)\beta^{-1}},
\end{equation*}
where $P$ is a constant and $\beta\geq1$ is the algebraic index of convergence.

Error $\varepsilon_{3}$ is the error related to the approximation of the Fourier cosine coefficients
of the density function in terms of its characteristic function, which reads
\begin{equation*}
|\varepsilon_{3}(x,N, [a,b])| = e^{-rT}\sum^{N-1}_{j=0}{^{'}}\mathrm{Re}\bigg[
\int_{\mathbb{R}\backslash[a,b]}e^{ik\pi\frac{y-a}{b-a}}e^{\alpha y}f_{T}(y|x)dy\bigg]V_{k}.
\end{equation*}
It can be shown that
\begin{equation*}
|\varepsilon_{3}(x,N, [a,b])| <e^{-rT}Q_{1} \int_{\mathbb{R}\backslash[a,b]}e^{\alpha y}f(y|x)dy,
\end{equation*}
where $Q_{1}$ is a constant independent of $N$ and $T$.

We denote by
\begin{equation*}
I_{1} = \int_{\mathbb{R}\backslash[a,b]}g(y)f_{T}(y|x)dy,\;\;\; I_{2} = \int_{\mathbb{R}\backslash[a,b]}e^{\alpha y}f_{T}(y|x)dy,
\end{equation*}
so that $\varepsilon_{1}= e^{-rT}I_{1}$, $\varepsilon_{3} < e^{-rT}Q_{1}I_{2}$. $\varepsilon_{3}$ can be controlled by $I_{2}$
Integral $I_{1}$ then depends on the payoff function
and the integration range, whereas $I_{2}$ depends only on the integration range.

For a call option, $g(y)=K(e^{y}-1)^{+}$, we have $\forall y$, $e^{-\alpha y}g(y)\leq Q_{2}(\alpha)$ when $\alpha>1$
where $Q_{2}(\alpha)$ depends on $\alpha$, so that it follows directly that
\begin{equation}\label{16}
I_{1}\leq Q_{2}(\alpha)I_{2},
\end{equation}
and $\varepsilon_{1}$ can be controlled by $I_{2}$ and $\alpha$. So overall errors are controlled by means of parameter $\alpha$,
$L$ and $N$.

Generally, for a call option, a large $\alpha$ reduces the cancellation errors of payoff function, but may lead to $I_{2}$ increase.
For a fixed $L$, when $f(y|x)$ has fat tails, $I_{2}$ may be dominated, so $\alpha$ must be small.

\subsection{The Analytic Solution for coefficient $V_{T}(u)$}

The coefficient $V_{T}(u)$ in (7) has analytic solution for several contracts.
In order to recover the coefficient $V_{T}(u)$, we first give following formulae
\begin{align}\label{17}
\chi(u,v;c,d)\equiv& \int^{d}_{c}e^{vy}\cos[u(y-a)]dy\nonumber\\
=&\frac{1}{v^{2}+u^{2}}\big\{-ve^{vc}\cos[u(c-a)]-ue^{vc}\sin[u(c-a)]\nonumber\\
&\hspace{15mm}+ve^{vd}\cos[u(d-a)]+ue^{vd}\sin[u(d-a)]\big\}
\end{align}

For European call, $g(y)=K(e^{y}-1)^{+}$, we have
\begin{align}\label{18}
V^{\mathrm{call}}_{T}(u)=&\frac{2}{b-a}\int^{b}_{a}e^{-\alpha y}K(e^{y}-1)^{+}\cos[u(y-a)]dy\nonumber\\
=&\frac{2K}{b-a}\big(\chi(u,1-\alpha;0,d)-\chi(u,-\alpha;0,d)\big).
\end{align}
Similarly, for European put, $g(y)=-K(e^{y}-1)^{+}$, we find
\begin{align}\label{19}
V^{\mathrm{put}}_{T}(u)=&-\frac{2}{b-a}\int^{b}_{a}e^{-\alpha y}K(e^{y}-1)^{+}\cos[u(y-a)]dy\nonumber\\
=&\frac{2K}{b-a}\big(-\chi(u,1-\alpha;a,0)+\chi(u,-\alpha;a,0)\big).
\end{align}

\section{Numerical Results}

In this section, we perform a variety of numerical
tests to evaluate the efficiency and accuracy of the Stable COS method.
The CPU used is an Intel(R) Core(TM) i7-6700 CPU (3.40GHz Cache size 8MB) with an implementation in Matlab 7.9.
Appendix contains Matlab code for implementing the Stable COS method to price European Call and Put options.

We focus on the plain vanilla European call options and consider different models for the underlying
asset from the the Heston stochastic volatility model, Kou jump-diffusion model, and CGMY model.

Table 2 presents the characteristic functions of $\ln(S_{t}/S_{0})$ for various models. 			
The parameters of various models for numerical experiment are given by Table 3.
In the CGMY model we choose $Y=1.5$ and 1.98 in the tests.

\vspace{2mm}

\begin{center}
Table 2: Characteristic functions of $\ln(S_{t}/S_{0})$ for various models.\\
{\footnotesize\begin{tabular}{r|l}
  \hline
Model&Characteristic function\\
  \hline
Heston&$\phi_{t}(u)=e^{A_{t}(u) + B_{t}(u) + C_{t}(u)}$\\
&$A_{t}(u)=iu(r-q)t$\\
&$B_{t}(u)=\frac{2\zeta(u)(1-e^{-\xi(u)t})V_{0}}{2\xi(u)-(\xi(u)-\gamma(u))(1-e^{-\xi(u)t})}$\\
&$C_{t}(u)=-\frac{\kappa\theta}{\sigma^{2}}\big[2\log\big(\frac{2\xi(u)-(\xi(u)-\gamma(u))(1-e^{-\xi(u)t})}{2\xi(u)}\big)+(\xi(u)-\gamma(u))t\big]$\\
&$\zeta(u)=-\frac{1}{2}(iu+u^{2})$\\
&$\xi(u)=\sqrt{\gamma(u)-2\sigma^{2}\zeta(u)}$\\
&$\gamma(u)=\kappa-i\rho\sigma u$\\
Kou& $\phi_{t}(u)=\exp\{iu\mu t-\frac{1}{2}\sigma^{2}u^{2}t+\lambda t\big(\frac{p\eta_{1}}{\eta_{1}-iu}+\frac{(1-p)\eta_{2}}{\eta_{2}+iu}-1\big)\big\}$\\
& $\mu = r-q-\frac{1}{2}\sigma^{2}-\lambda \big(\frac{p\eta_{1}}{\eta_{1}-1}+\frac{q\eta_{2}}{\eta_{2}+1}-1\big)$\\
CGMY& $\phi_{t}(u)=e^{iu\mu t}\exp\{Ct\Gamma(-Y)[(M-iu)^{Y}-M^{Y}+(G+iu)^{Y}-G^{Y}]\}$\\
&$\mu=r-q-C\Gamma(-Y)\big((M-1)^{Y}-M^{Y}+(G+1)^{Y}-G^{Y}\big)$\\
  \hline
\end{tabular}}
\end{center}

\begin{center}
Table 3: Model parameters of various models in numerical experiment\\
{\footnotesize\begin{tabular}{l|l}
  \hline
Common for all Models& $S_{0} = 100$, $r = 0.1$, $q = 0$\\
  \hline
Model&  parameters\\
   \hline
Heston&$\kappa=0.85$, $\theta=0.30^{2}$, $\sigma=0.1$, $\rho=-0.7$, $V_{0}=0.25^2$\\
Kou&$\sigma=0.16$, $p= 0.4$, $\eta_1 = 10$, $\eta_2 = 5$, $\lambda = 5$\\
CGMY\_1&$C = 1$, $G = 5$, $M = 5$, $Y = 1.5$\\
CGMY\_2&$C = 1$, $G = 5$, $M = 5$, $Y = 1.98$\\
  \hline
\end{tabular}}
\end{center}

\begin{figure}[h]
  \vspace{0mm}
\begin{center}
  \includegraphics[width=10cm]{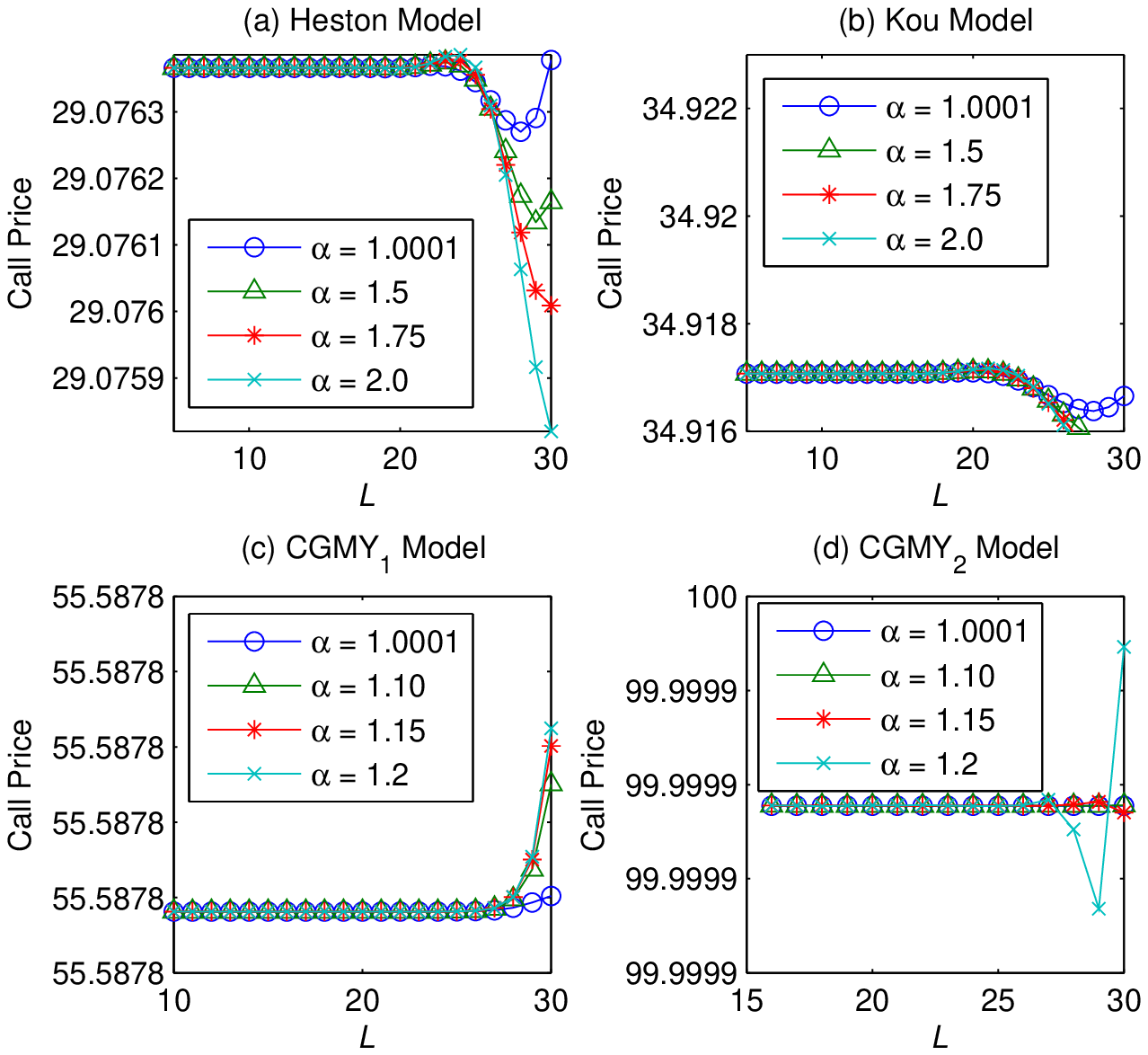}\\
  \vspace{-5mm}
  \caption{Damping parameter $\alpha$ and truncation parameter $L$ for Stable\_Cos method.}\label{1}
\end{center}
\end{figure}

\begin{figure}[h]
  \vspace{-3mm}
\begin{center}
  \includegraphics[width=10cm]{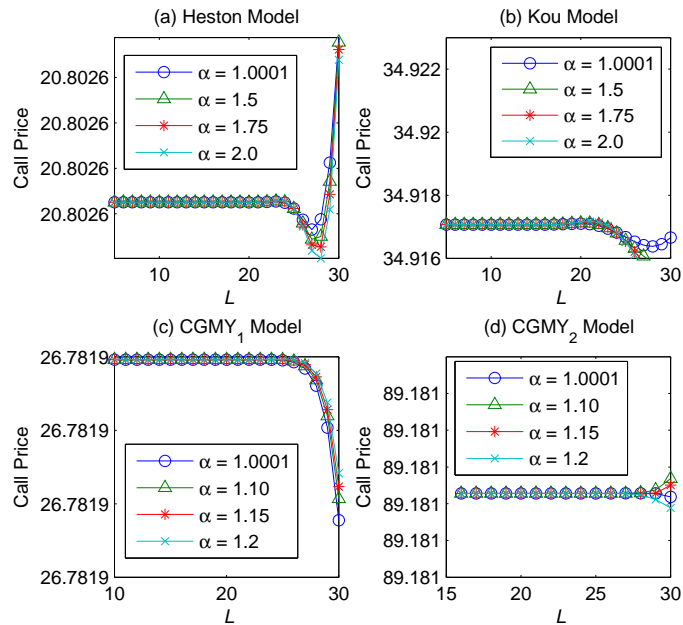}\\
  \vspace{-4mm}
  \caption{Damping parameter $\alpha$ and truncation parameter $L$ for Stable Cos method when $T=0.1$.}\label{2}
\end{center}
    \vspace{-4mm}
\end{figure}

We compare our results with the Stable COS methods to two of Cos method,
the direct Cos method and put-call parity Cos method in which the put price is
calculated first, then by put-call parity, call price is obtained.
We reference Stable\_Cos to Stable COS method,
Put-Call\_Cos to put-call parity Cos method, and Direct\_Cos to the direct Cos method. We have three kinds Cos methods.

\subsection{Damping Factors and Truncation Range}

In this subsection we consider the choice of the damping parameter $\alpha$ and truncation interval $[a,b]$.
In order to illustrate the result numerically, we have chosen different values of $\alpha$ and
$K = 80$, $T = 1$ for all models considered in this paper to
generate the graphs given in Figure 1 by Stable Cos method.
The reference value for the European option can be found from Table 5.

Figures 1 presents European call option values under different damping parameters $\alpha$ and range of Truncation parameters $L$.
In Figure 1, the option values obtained by Stable Cos method.
Figures 1 shows that option values are stable under $\alpha\in[1.0001,1.2]$ for all cases by Stable Cos method,
and for most cases, $L\in[6,18]$ is reasonable except that the probability density function of the underlying is governed by fat tails.
For fat tail cases, option values are stable under $L\in[17,25]$.
Figures 3 and 4 show that such results are also robust for different $T$-values.

\begin{figure}[h]
  \vspace{0mm}
\begin{center}
\begin{center}
  \includegraphics[width=15cm]{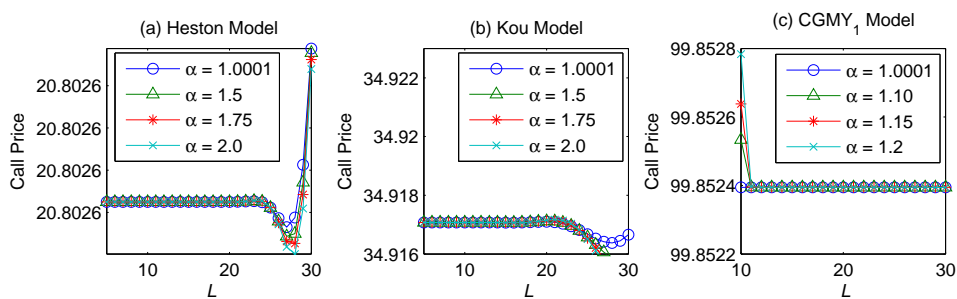}\\
\end{center}
  \vspace{-4mm}
  \caption{Damping parameter $\alpha$ and truncation parameter $L$ for Stable Cos method when $T=20$.}\label{3}
    \vspace{-4mm}
\end{center}
    \vspace{-4mm}
\end{figure}

\subsection{accuracy, efficiency and robustness of R\_Cos}

Now we examine the accuracy, efficiency and robustness of our robust Cos methods by a
series of numerical examples. For further comparison, we use Carr-Madan method (Carr and Madan (1999))
to calculate call price for various models. In case of the Carr-Madan method, we use FFT method to calculate call prices with
grid points $N=2^{16}$ and damping factor $\alpha=0.75$, and apply cubic interpolation to obtain desirable price.
Moreover, we also use Fourier transform method (FTM) (Eberlein, Glau and Papapantoleon (2010)) to calculate call price.
In later case, we use matlab built-in function \texttt{quadgk} to calculate the integrals with
integral interval $[-5000,5000]$ and damping factor $\alpha=1.1$ for Heston, Kou,
and CGMY\_1 and damping factor $\alpha=1.015$ for CGMY\_2.

In the experiments, the parameters of Cos methods is given by Table 4 where the values of $\alpha$, $L$ and $N$ are included for various models.
These parameters are chosen such that same accuracy is obtained as as possible.
Table 5 presents values of European call option to round ten decimals for a series of strike prices with $T=1$ using Stable Cos method,
Put-call\_Cos method, direct Cos method, FFT method and Fourier transform method.
From Table 5, we can get the impression that the same accuracy is obtained by Stable Cos method and Put-call Cos method.

\begin{center}
Table 4 The method parameters for calculating\\ call price by Three kinds of Cos methods\\
{\footnotesize\begin{tabular}{l|ccc|cc|cc}
  \hline
&  \multicolumn{3}{c|}{Stable\_Cos}&\multicolumn{2}{c|}{Put-Call\_Cos}&\multicolumn{2}{c}{Direct\_Cos}\\
\cline{2-8}
Model &Damping& $L$ & $N$ &$L$ & $N$ & $L$ & $N$\\
  \hline
Heston & 1.1  & 7 & 110&   7& 110& 7 &   110\\
Kou    & 1.1  & 7 & 140&   11& 210& 10 &  210\\
CGMY\_1& 1.001& 10& 50 &   10&  50&  13 &  80\\
CGMY\_2& 1.001& 17& 80 &   10&  70& &  \\
  \hline
\end{tabular}}
\end{center}

For efficiency comparison, we calculate the absolute errors of values of call option for a series of $N$ using four kinds of Cos methods with $K=100$ and $T=1$.
The other method parameters is given by Table 4 and the reference values is given in Table 6.
The computing results are plotted in Figure 4. As shown in Figure 4, the error convergence of Stable\_Cos method is same as
or superior to that of Put-Call\_Cos method except CGMY\_2,
where error convergence of Stable\_Cos is sightly inferior to that of Put-Call\_Cos but still exponential.

The convergence results are not sensitive for different $T$-values. Figure 5 and 6 present error convergence results for CGMY\_1 with $T=5$
and CGMY\_2 with $T=0.1$ by Stable\_Cos and Put-Call\_Cos methods. As shown in Figure 5 and 6, error convergence results do not change much as $T$ changes.

\begin{center}
Table 5 Values of Option Price in Various Models with $T=1$\\
{\footnotesize\tabcolsep0.05in\begin{tabular}{rccccc}
  \hline
  \hline
Strike& Stable\_Cos& Put-Call\_Cos & Direct\_Cos& FTM & FFT\\
    \hline
\multicolumn{6}{c}{Heston Model}\\
  \hline
80&	29.0763658809& 	29.0763658809& 	29.0763658809& 	29.0763658809& 	29.0761596018\\
85&	25.3190242256& 	25.3190242256& 	25.3190242256& 	25.3190242256& 	25.3182564563\\
90&	21.8125703138& 	21.8125703138& 	21.8125703138& 	21.8125703138& 	21.8118590739\\
95&	18.5866087601& 	18.5866087601& 	18.5866087601& 	18.5866087601& 	18.5863467950\\
100&15.6621055646& 	15.6621055646& 	15.6621055646& 	15.6621055646& 	15.6621055646\\
105&13.0502592738& 	13.0502592738& 	13.0502592738& 	13.0502592738& 	13.0499484651\\
110&10.7523654075& 	10.7523654075& 	10.7523654075& 	10.7523654075& 	10.7513381775\\
115& 8.7605509099& 	 8.7605509099& 	 8.7605509099& 	 8.7605509099& 	 8.7590733973\\
120& 7.0591610639& 	 7.0591610639& 	 7.0591610639& 	 7.0591610639& 	 7.0581882812\\
  \hline
\multicolumn{6}{c}{Kou Model}\\
\hline
80&	34.9170704483& 	34.9170704483& 	34.9170704483& 	34.9170704483& 	34.9170564801\\
85&	31.9123200707& 	31.9123200707& 	31.9123200707& 	31.9123200707& 	31.9123052075\\
90&	29.0794136987& 	29.0794136987& 	29.0794136987& 	29.0794136987& 	29.0793943989\\
95&	26.4197703718& 	26.4197703718& 	26.4197703718& 	26.4197703718& 	26.4197774446\\
100&23.9335400091& 	23.9335400091& 	23.9335400091& 	23.9335400091& 	23.9335400091\\
105&21.6196765651& 	21.6196765651& 	21.6196765651& 	21.6196765651& 	21.6196851589\\
110&19.4760004471& 	19.4760004471& 	19.4760004471& 	19.4760004471& 	19.4759675797\\
115&17.4992412865& 	17.4992412865& 	17.4992412865& 	17.4992412865& 	17.4992369744\\
120&15.6850612076& 	15.6850612076& 	15.6850612077& 	15.6850612076& 	15.6850931891\\
    \hline
\multicolumn{6}{c}{CGMY\_1}\\
\hline
80&	55.5877500641& 	55.5877500641& 	55.5877500089& 	55.5877500641& 	55.5877433925\\
85&	54.0282287092& 	54.0282287092& 	54.0282286548& 	54.0282287092& 	54.0281877377\\
90&	52.5459973200& 	52.5459973200& 	52.5459972649& 	52.5459973200& 	52.5459627032\\
95&	51.1352855665& 	51.1352855665& 	51.1352855100& 	51.1352855665& 	51.1352741113\\
100&49.7909054685& 	49.7909054685& 	49.7909054141& 	49.7909054685& 	49.7909054685\\
105&48.5081777104& 	48.5081777104& 	48.5081776544& 	48.5081777104& 	48.5081662419\\
110&47.2828690189& 	47.2828690189& 	47.2828689625& 	47.2828690189& 	47.2828329497\\
115&46.1111387169& 	46.1111387169& 	46.1111386614& 	46.1111387169& 	46.1110877016\\
120&44.9894929189& 	44.9894929189& 	44.9894928638& 	44.9894929189& 	44.9894576019\\
    \hline
\multicolumn{6}{c}{CGMY\_2}\\
  \hline
80&	99.9999155240& 	99.9999155240&& 		99.9999155240& 	99.9999155240\\
85&	99.9999129093& 	99.9999129092&& 		99.9999129092& 	99.9999129093\\
90&	99.9999103728& 	99.9999103728&& 		99.9999103728& 	99.9999103728\\
95&	99.9999079083& 	99.9999079083&& 		99.9999079082& 	99.9999079083\\
100&99.9999055101& 	99.9999055101&& 		99.9999055100& 	99.9999055101\\
105&99.9999031733& 	99.9999031732&& 		99.9999031732& 	99.9999031733\\
110&99.9999008935& 	99.9999008935&& 		99.9999008935& 	99.9999008935\\
115&99.9998986670& 	99.9998986669&& 		99.9998986669& 	99.9998986670\\
120&99.9998964902& 	99.9998964902&& 		99.9998964901& 	99.9998964902\\
  \hline
\end{tabular}}
\end{center}
\vspace{-2mm}
\hspace{10mm}{\scriptsize Note: FTM reference to Fourier transform method (Eberlein, Glau and Papapantoleon (2010)).}

\vspace{4mm}

Finally, we consider the robustness of our methods. For given the method parameters in Table 4, we calculate values of call price by Stable\_Cos method
and Put-Call\_Cos method for range of $L$-values $L$. The computing results are shown in Figure 7. From Figure 7, we find that
size of the integration interval is almost same for two methods, so our method has same robustness as Put-Call\_Cos method.

\begin{figure}
  \vspace{0mm}
\begin{center}
  \includegraphics[width=10cm]{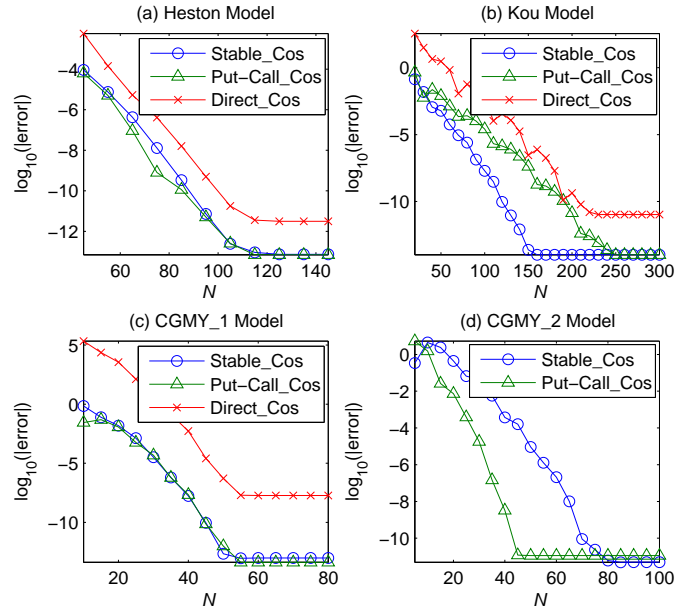}\\
  \vspace{-4mm}
  \caption{Convergence of Stable Cos method, Put-Call Cos method and Direct Cos method.}\label{4}
\end{center}
  \vspace{-4mm}
\end{figure}

\begin{figure}[ht]
  \vspace{0mm}
\begin{center}
  \includegraphics[width=15cm]{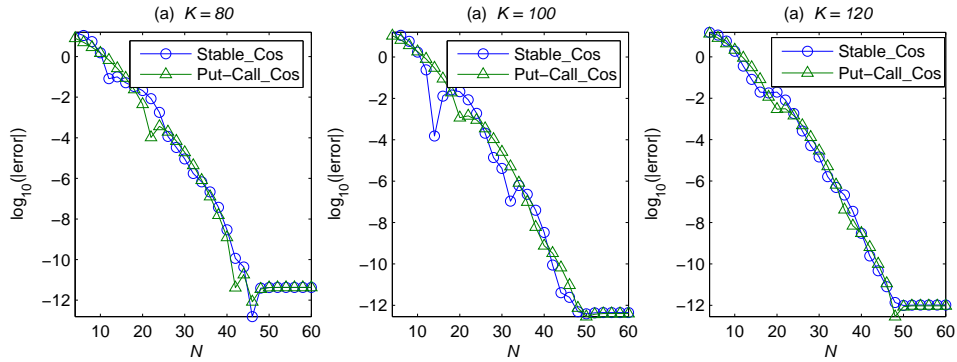}\\
  \vspace{0mm}
  \caption{Convergence of Stable\_Cos method and Put-Call\_Cos method for CGMY\_1 with $T=5$.}\label{5}
\end{center}
\end{figure}

\newpage

\begin{center}
Table 6 The reference values for calculate the absolute errors\\ for four kinds of Cos methods with $K=100$ and $T=1$\\
\tabcolsep0.8in\begin{tabular}{l|l}
  \hline
Model & reference values\\
  \hline
Heston & 15.6621055645751\\
Kou&     23.9335400090856\\
CGMY\_1& 49.7909054685239\\
CGMY\_2& 99.9999055100654\\
  \hline
\end{tabular}
\end{center}
\vspace{-2mm}
\hspace{10mm}
\begin{minipage}[l]{130mm}
    {\footnotesize Note: The reference values are obtained by Put-Call\_Cos with $N=60000$.}
\end{minipage}

\vspace{2mm}

\begin{figure}[h]
  \vspace{0mm}
\begin{center}
  \includegraphics[width=15cm]{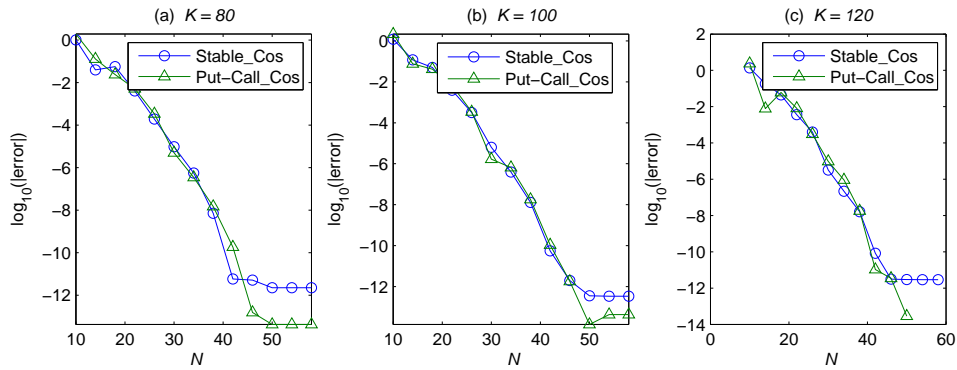}\\
  \vspace{-4mm}
  \caption{Error convergence of Stable\_Cos method and Put-Call\_Cos method for CGMY\_2 with $T=0.1$.}\label{6}
    \vspace{-4mm}
\end{center}
\end{figure}

\begin{figure}[ht]
  \vspace{0mm}
\begin{center}
  \includegraphics[width=15cm]{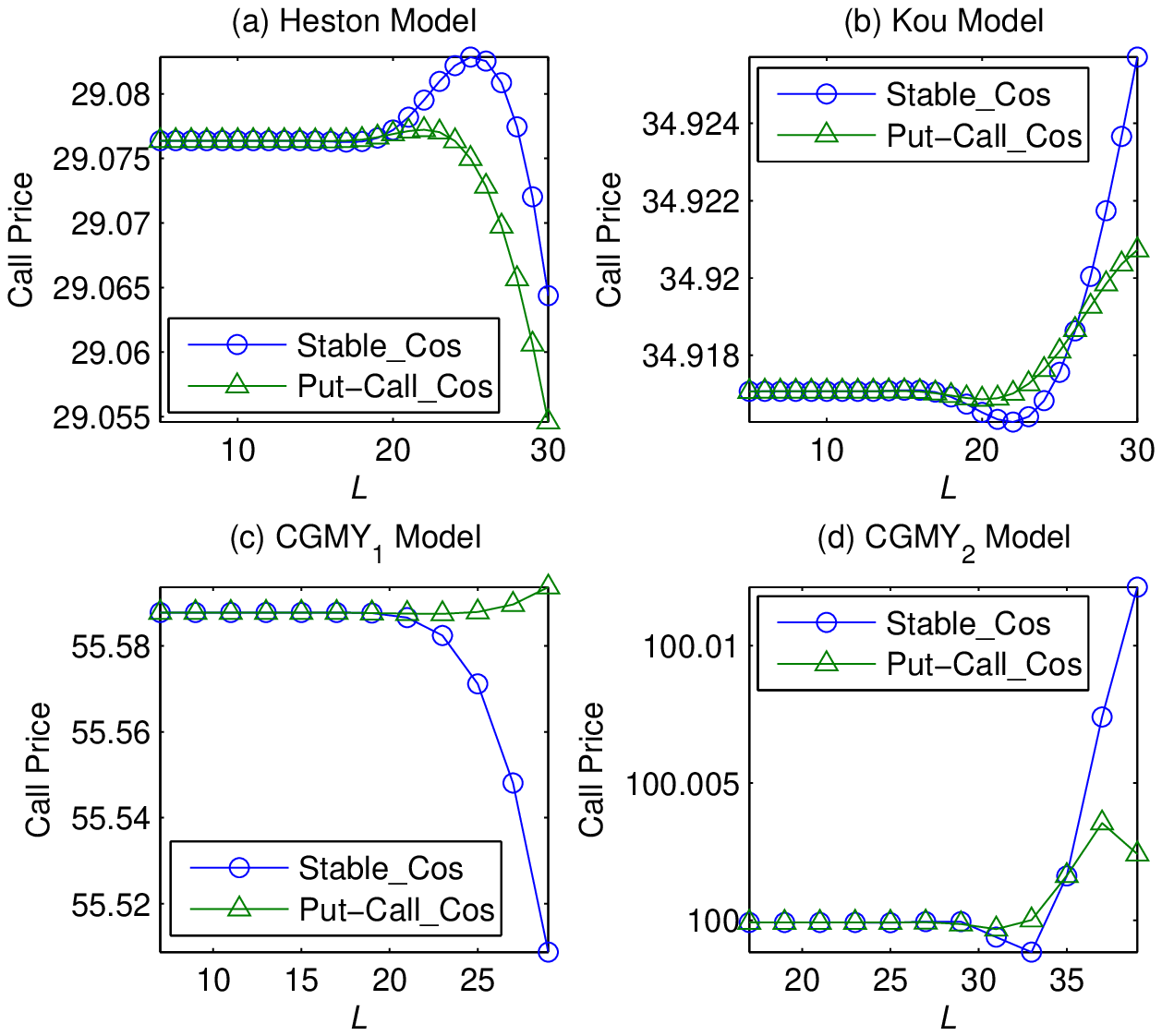}\\
  \vspace{-4mm}
  \caption{Comparison of $L$-values by Stable\_Cos method and Put-Call\_Cos method.}\label{7}
\end{center}
\end{figure}

\section{Conclusions}

In this paper, we present a robust pricing of call options based on Fourier cosine series expansion.
The robust COS method exhibits an exponential convergence in $N$ for density functions in $C^{\infty}[a,b]$ and an impressive
computational speed. With a limited number, $N$, of Fourier cosine coefficients,
it produces highly accurate results. We also present error analysis for this
method, showing that error convergence is easily obtained. Robust pricing, insensitive of the choice of the size of the
integration range is achieved for call options.
The accuracy, efficiency and robustness of our robust Cos methods
are demonstrated by error analysis, as well as by a series of numerical examples, including
Heston stochastic volatility model, Kou jump-diffusion model,
and CGMY model.

\end{document}